\documentclass[]{spie}  

\usepackage{amsmath,amsfonts,amssymb}
\usepackage{graphicx}
\usepackage{comment}
\usepackage[colorlinks=true, allcolors=blue]{hyperref}

\title{What could KIDSpec, a new MKID spectrograph, do on the ELT?}

\author[a]{V. Benedict Hofmann}
\author[a]{Kieran O'Brien}
\author[a]{Deli Geng}
\affil[a]{Centre for Advanced Instrumentation, Department of Physics, Durham University, South Road, Durham, DH1 3LE, UK}

\authorinfo{Further author information: (Send correspondence to V.B.H.)\\V.B.H.: E-mail: benedict.hofmann@durham.ac.uk}

\begin{document} 
\maketitle

\begin{abstract}
Microwave Kinetic Inductance Detectors (MKIDs) are beginning to become more prominent in astronomical instrumentation, due to their sensitivity, low noise, high pixel count for superconducting detectors, and inherent energy and time resolving capability. The Kinetic Inductance Detector Spectrometer (KIDSpec) will take advantage of these features, KIDSpec is a medium resolution MKID spectrograph for the optical/near infrared. KIDSpec will contribute to many science areas particularly those involving short and/or faint observations. When short period binary systems are found, typical CCD detectors will struggle to characterise these systems due to the very short exposures required, causing errors as large as the estimated parameter itself. The KIDSpec Simulator (KSIM) has been developed to investigate how much KIDSpec could improve on this. KIDSpec was simulated on an ELT class telescope to find the extent of its potential, and it was found that KIDSpec could observe a $m_{V}\approx{24}$ with an SNR of 5 for a 10s exposure at 1420 spectral resolution. This would mean that KIDSpec on an ELT class telescope could spectroscopically follow up on any LSST photometric discoveries of LISA verification sources.
\end{abstract}

\keywords{Instrument simulation, microwave kinetic inductance detectors, spectroscopy}

\section{INTRODUCTION}
\label{sec:intro}  

\subsection{Microwave Kinetic Inductance Detectors and KIDSpec}
\label{sec: MKIDs}

Microwave Kinetic Inductance Detectors (MKIDs) are low-temperature superconducting detecctors, and in the optical (OPT)/near-infrared (NIR) continue to grow in interest as they are used in and suggested for various instruments. Some examples include the Array Camera for Optical to Near-IR Spectrophotometry (ARCONS)\cite{Mazin2013ARCONS:Spectrophotometer}, the DARK-speckle Near-infrared Energy-resolving Superconducting Spectrophotometer (DARKNESS)\cite{Meeker2018DARKNESS:Astronomy}, and the MKID Exoplanet Camera (MEC)\cite{Walter2020TheSCExAO}. MKID technology has also seen use in other wavelength ranges such as sub-mm\cite{Wilson2020TheResults,Brien2018MUSCAT:AsTronomy} . MKIDs are an exciting detector in the OPT/NIR due to being free of read noise and dark current counts, energy resolving, time resolving, and photon counting detectors\cite{OBrien2020KIDSpec:Spectrograph}. In particular by using the energy resolving ability of the MKIDs, they can be used as a cross-disperser in a spectrograph, due to them being able to separate the orders falling on a linear array of MKIDs.  

Taking advantage of this will be the Kinetic Inductance Detector Spectrometer (KIDSpec), an upcoming medium spectral resolution OPT/NIR spectrometer\cite{OBrien2020KIDSpec:Spectrograph}. While intended to have similar capabilities to X-Shooter\cite{Vernet2011X-shooterTelescope} , KIDSpec will instead use MKIDs as its detectors, gaining the benefits described above. The MKIDs function as a resonant circuit made up of a capacitor and inductor. Incoming photons break Cooper pairs in the superconducting material generating quasiparticles. The energy of this incoming photon then determines how many of these Cooper pairs are broken, due to the energy required to break a Cooper pair being much lower than photons in the OPT/NIR\cite{OBrien2014KIDSpec:Spectrograph}. As a result of this quasiparticle generation the surface impedance of the superconductor changes. To detect these changes in the MKID, a microwave signal can be tuned to the resonant frequency of the MKID resonator. The change in surface impedance then causes a particular shift in the phase and amplitude of this microwave signal, which can be measured. By tracking the phase of the microwave signal with time, incoming photons can be seen as a fast rise and exponential decay, to the order of microseconds time resolution. The height of these fast rises is the result of the number of quasiparticles generated from the number of Cooper pairs broken, and as such the energy of the incoming photon. An important feature for KIDSpec is the energy resolution ($R_{E}$) of its MKIDs. If a monochromatic source was exposed onto an MKID, the MKID measures a range of wavelengths for this source. This distribution appears as a Gaussian\cite{Meeker2015DesignImaging} and the $R_{E}$ is the FWHM of this Gaussian. Each MKID in KIDSpec will observe a wavelength from each grating order. As a result of the energy resolution each order response by the MKID post-observation will have the approximate form of a Gaussian. To separate the orders with a $3\sigma$ separation, the number of spectral orders from a grating that can be separated by KIDSpec can be defined by $R_{E} \geq 3m$, where $m$ is the order number\cite{Hofmann2022KSIMNIR} . Recently an $R_{E}$ of $\approx{55}$ at 402nm\cite{deVisser2020ExperimentalMKIDs} has been demonstrated. For examples of phase-time streams and more details on MKIDs see \cite{Day2003AArrays,Meeker2015DesignImaging,Mazin2004MicrowaveDetectors,Mazin2019Optical2020s} . 

\subsection{Short Period Binary systems}
\label{sec: binaries}
The Vera Rubin Observatory is predicted to detect tens of millions of eclipsing binary systems during its survey, and Gaia was predicted to observe 30 million non-single stars and 8 million spectroscopic binaries. So when short period systems are found, typical CCD detectors will struggle to characterise these systems due to the very short exposures required, causing errors as large as the estimated parameter itself\cite{Burdge2019GeneralSystem} . These systems are also of great interest as they are LISA verification sources \cite{Kupfer2018LISA2,Amaro-Seoane2017LaserProposal} . It was predicted that LSST would be able to find these systems up to a magnitude of $\approx{24}$\cite{Korol2017ProspectsLISA} , but spectroscopic follow up will be non-trivial with typical CCDs. KIDSpec however, using its low noise detectors present a solution to this issue.

\section{KIDSpec Design and Simulation}

   \begin{figure} [ht]
   \begin{center}
   \begin{tabular}{c}  
   \includegraphics[height=7cm]{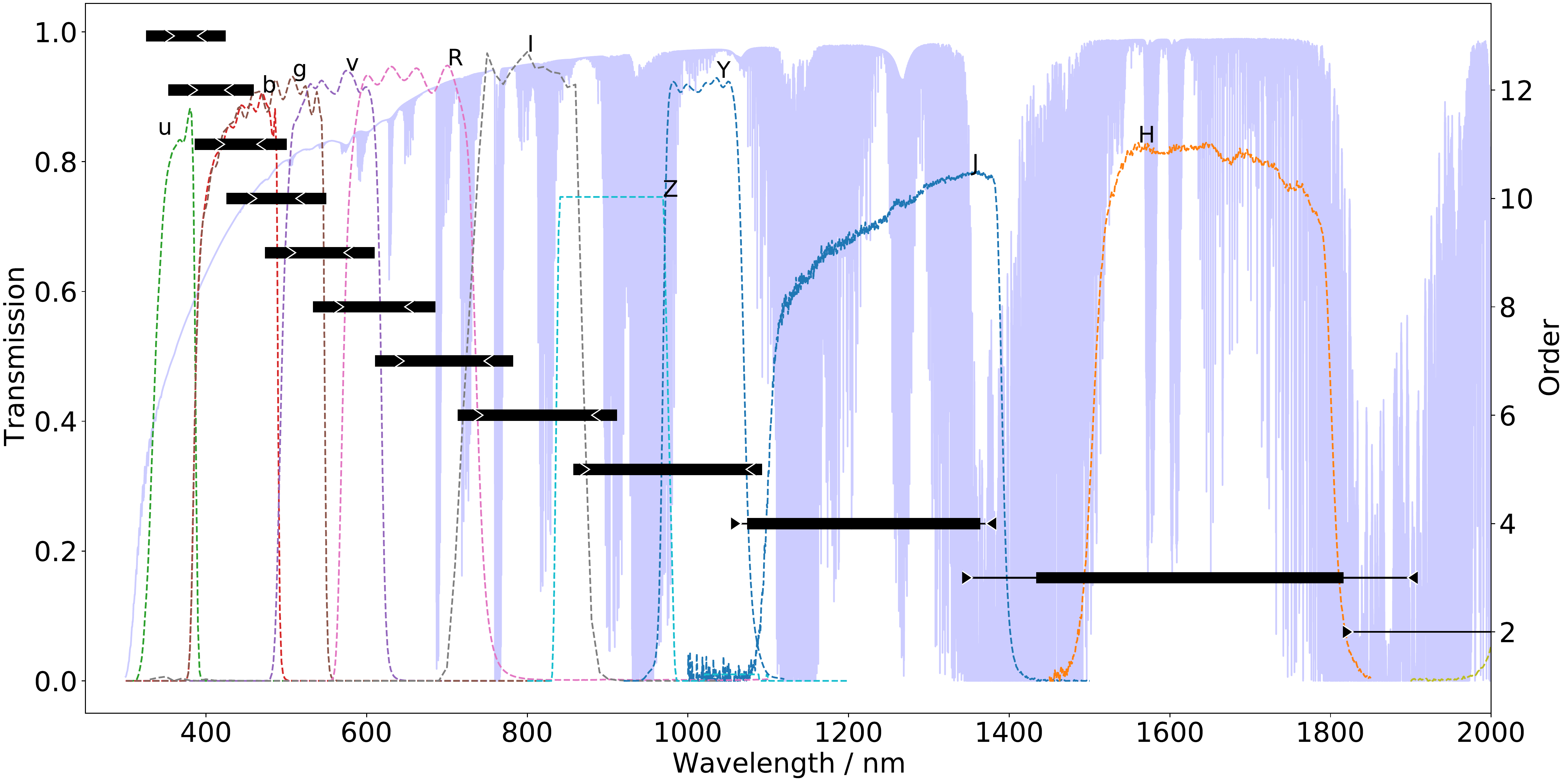}
   \end{tabular}
   \end{center}
   \caption[example] 
   { \label{fig:7500_orders} 
Optimiser grating parameters for a $R_{E}$ of 40 with spectral resolution 7500. Plotted are the grating order wavelength ranges observed by the MKIDs in bold black bars. The free spectral range for each order is represented by the thinner black lines and arrows pointing inwards. Magnitude bands from ESO used for ETC simulations are also plotted, with the GEMINI atmospheric transmission data. This gave a central first order wavelength of $4.9\mu m$ with 3400 MKIDs in the linear array. In the region of $\approx{1.4}\mu m$, the optimiser has excluded this region due to the poor atmospheric transmission.}
   \end{figure} 

To determine a KIDSpec design for this, an optimiser was used to assign grating parameters for the instrument. This optimiser takes in a desired $R_{E}$ and spectral resolution to find the most suitable grating order placements with respect to wavelength. Accounted for is the atmospheric transmission and sky brightness. Atmospheric data is used from GEMINI\footnote{\url{http://www.gemini.edu/observing/telescopes-and-sites/telescopes}} and the ESO SKYCALC Sky Model Calculator\footnote{\url{http://www.eso.org/observing/etc/bin/gen/form?INS.MODE=swspectr+INS.NAME=SKYCALC}} is used for the sky. A $R_{E}$ of 40 was chosen as this would allow enough orders to be separated for a spectral resolution of 7500, which is high enough to track spectral lines in a variety of astrophysical objects\cite{DOdorico2004X-shooter:VLT} , particularly for an instrument with a wide bandpass such as KIDSpec. The result of the optimiser for the grating orders is shown in Fig. \ref{fig:7500_orders}, giving a central first order wavelength of $4.9\mu m$ with 3400 MKIDs in the linear array. This design for KIDSpec was then simulated in this work using the KIDSpec Simulator (KSIM), a end-to-end simulator to estimate how KIDSpec designs could perform on sky. More details on the optimiser and KSIM can be found in \cite{Hofmann2022KSIMNIR} .

\section{Simulations}
\label{sec: sims}

\subsection{8m telescope}
\label{sec: 8m_sims}
\begin{figure} [ht]
   \begin{center}
   \begin{tabular}{c}  
   \includegraphics[height=8cm]{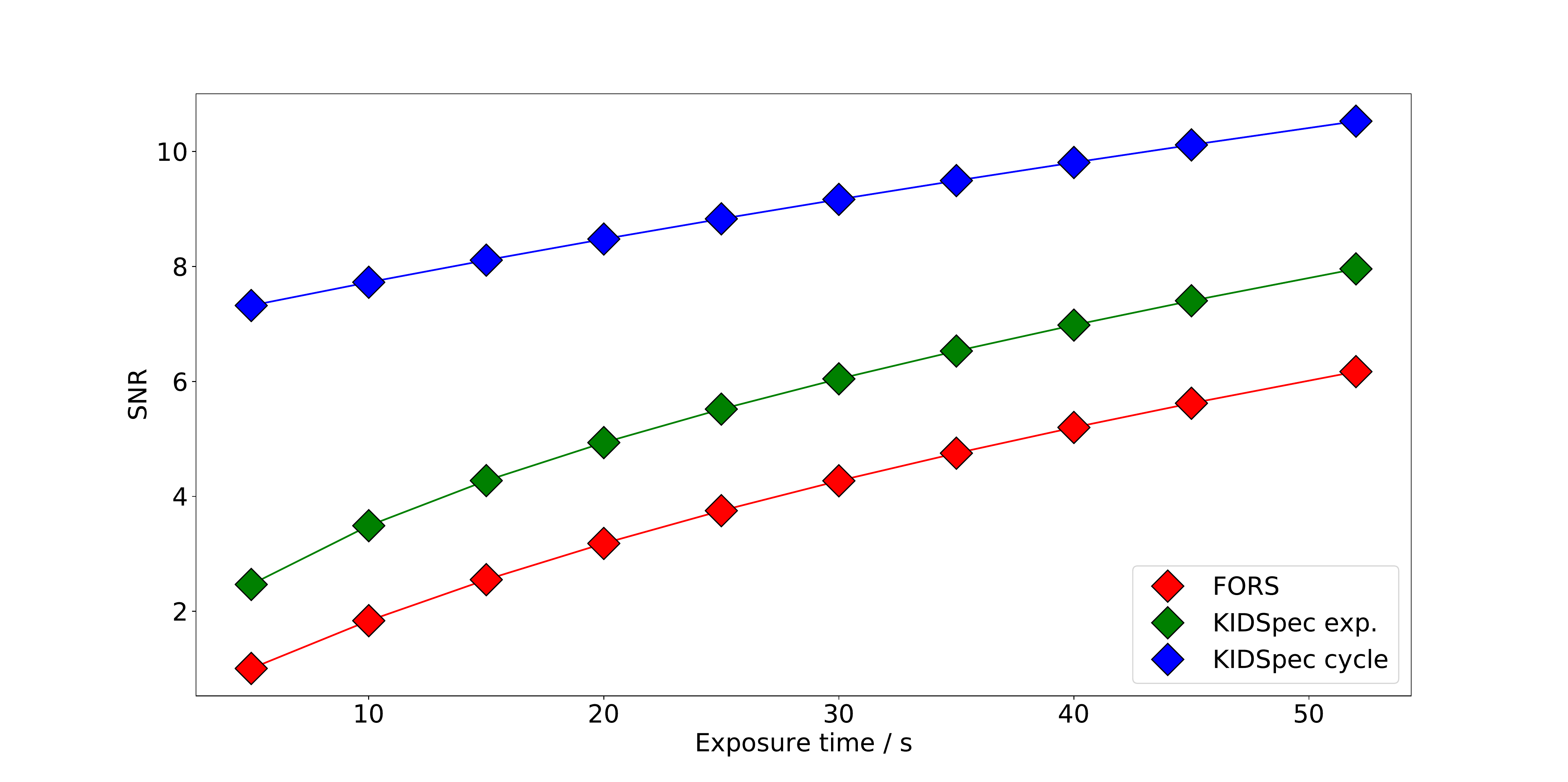}
   \end{tabular}
   \end{center}
   \caption[example] 
   { \label{fig:fors_comp} 
SNR results at 452nm for simulating a ZTFJ1539 + 5027 like system at various exposures. Also simulated were the cycle times for these FORS exposures. SNR taken at a spectral resolution of 1420, to align with the $\approx{1600}$ spectral resolution setup used by LRIS for the original observations of ZTFJ1539 + 5027.}
   \end{figure}

To demonstrate the gains KIDSpec could make over conventional detectors, its performance using KSIM was compared to FORS' using its ESO ETC. Both were simulated on an 8m diameter telescope. Simulated was a spectrum of ZTFJ1539 + 5027 like system, the binary observed in \cite{Burdge2019GeneralSystem} . This system has a magnitude of $\approx{19}$ and a period of 7 minutes. The setup chosen for FORS was a spectral resolution of 1420, due to this being closest to the $\approx{1600}$ spectral resolution used for the original observations with LRIS. Through the lack of read noise of the MKIDs, it is possible for KIDSpec's original resolution of 7500 exposure to be flexibly rebinned down to 1420, without requiring different gratings or setups. Additionally since KIDSpec would not need to be readout like a typical CCD detector, this would give more time on sky observing the object rather than reading out during a night on a telescope. This is considered here, with the setup used for FORS this would require a readout time of 39s\footnote{\url{https://www.eso.org/sci/facilities/paranal/instruments/fors/doc/VLT-MAN-ESO-13100-1543_P07.pdf}}. Here this would mean KIDSpec would be observing for an extra 39s on top of the original exposure time, the total hereafter in this paper named the cycle time. The ZTFJ1539 + 5027 like system was simulated for both instruments at varying exposure times, and the SNR at FORS' central wavelength of 452nm was taken. The same wavelength bin was used for the KIDSpec result. Results are shown in Fig. \ref{fig:fors_comp}.

From Fig. \ref{fig:fors_comp} at the shorter exposures such as 10s, KIDSpec at an SNR of 3.5 almost doubles the SNR of FORS at 1.8. This is as expected due to KIDSpec's MKIDs not suffering from read noise and dark current counts. When the cycle time is considered, for a FORS exposure of 10s at 1.8 SNR KIDSpec had an SNR of 7.7, a four times increase over the FORS result. As the exposure time increases the gains KIDSpec makes reduces, down to 1.3 times the FORS SNR of 6.2. This is expected as the exposure time increases since the impact of the read noise reduces with longer exposures where you are no longer read noise dominated.

\subsection{40m telescope}
\label{sec: 40m_sims}

Since there are gains to made using KIDSpec with its MKIDs, the potential of KIDSpec on an ELT class telescope of diameter 39.3m was simulated using KSIM. The ELT's transmission for M1-6 were considered, and parameters for the atmospheric conditions from ESO were also used\cite{Clenet2015AnisoplanatismField} . The ZTFJ1539 + 5027\cite{Burdge2019GeneralSystem} like system was simulated here with reduced magnitudes to find the limit for KIDSpec on the ELT. The threshold for this limit was 5, as this was the likely target for the original LRIS observations, since the FORS ETC simulations resulted in SNRs of $\approx{5}$ when replicating the original observations in Burdge \textit{et al}\cite{Burdge2019GeneralSystem} . The SNR was taken in the same way as Sec. \ref{sec: 8m_sims}. Also as in Sec. \ref{sec: 8m_sims}, the KSIM 7500 spectral resolution result was rebinned to 1420. KIDSpec using its MKIDs would not be limited by requiring a readout, so in addition to using the MKIDs time resolution, signal in a particular time bin can be built up with a continuous long observation. In Burdge \textit{et al}\cite{Burdge2019GeneralSystem} 317 52s exposures were used, so with readout time this may have been a nights worth of observations. Several observation times for this object were simulated; one 10s time bin in the period as done in Sec. \ref{sec: 8m_sims}, one hour of observation, and 10 hours of observation. 

   \begin{figure} [ht]
   \begin{center}
   \begin{tabular}{c}  
   \includegraphics[height=8cm]{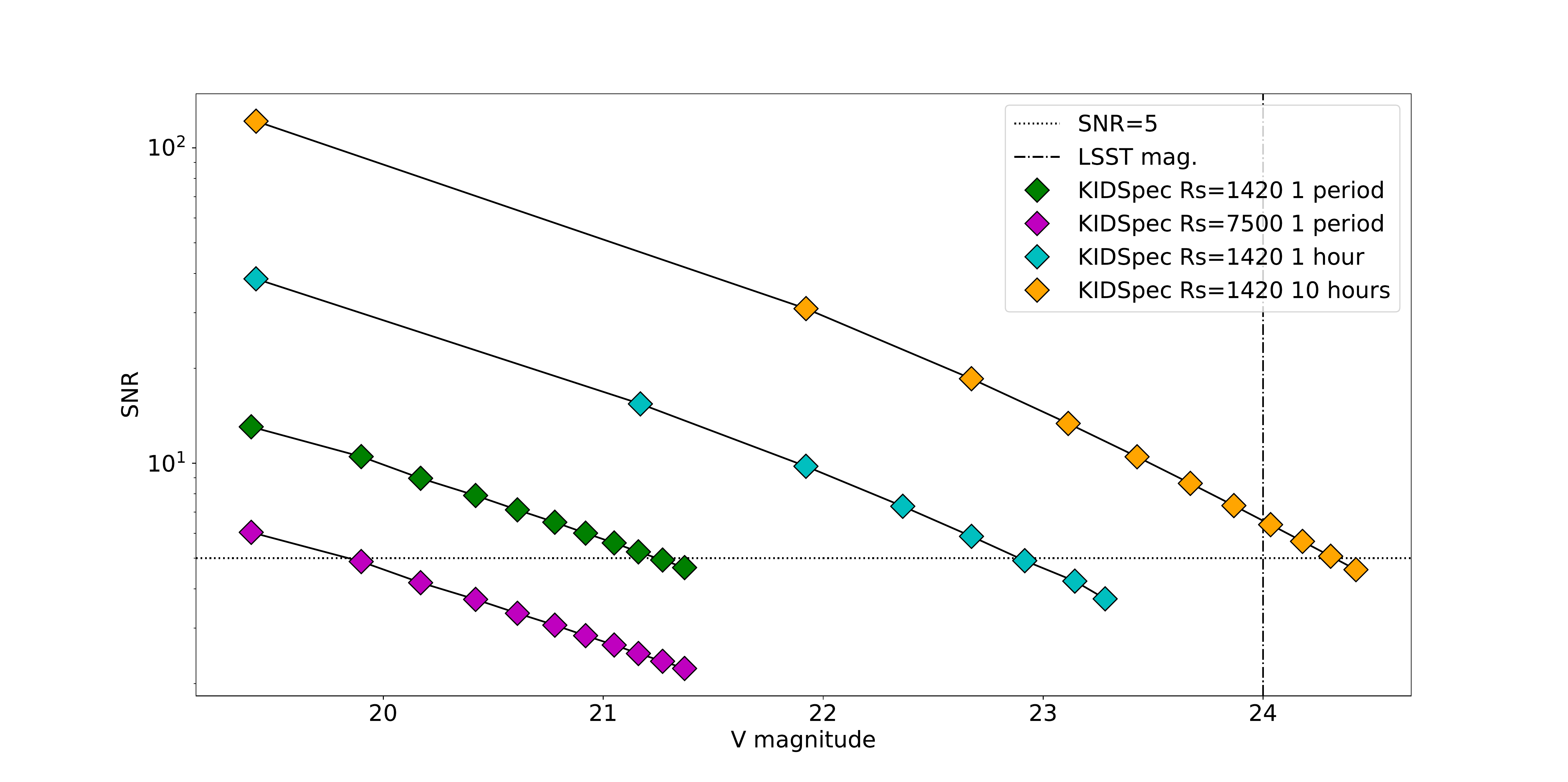}
   \end{tabular}
   \end{center}
   \caption[example] 
   { \label{fig:elt_mags} 
Magnitude limits for various exposures observing ZTFJ1539 + 5027 with an SNR of 5. With a spectral resolution of 7500, KIDSpec could potentially reach $m_{V}\approx{19.9}$ for one period which the approximate magnitude of ZTFJ1539 + 5027. When rebinned to 1420 resolution this increases to $m_{V}\approx{21.3}$. Extending this to an hour of observation would allow KIDSpec to reach $m_{V}\approx{22.9}$, and finally with 10 hours achieving $m_{V}\approx{24.3}$. This passes the magnitude of 24 threshold from LSST for these LISA verification sources.}
   \end{figure} 

Fig. \ref{fig:elt_mags} contains the results for the magnitude limits of these exposures. With a spectral resolution of 7500, KIDSpec could potentially reach $m_{V}\approx{19.9}$ for one orbital period which the approximate magnitude of ZTFJ1539 + 5027. When rebinned to 1420 resolution this increases to $m_{V}\approx{21.3}$. Extending this to an hour of observation would allow KIDSpec to reach $m_{V}\approx{22.9}$, and finally with 10 hours achieving $m_{V}\approx{24.3}$. With a night of observation then on a 40m diameter telescope KIDSpec could spectroscopically follow up on any LISA verification sources LSST photometrically finds in the sky. 

\begin{figure} [ht]
   \begin{center}
   \begin{tabular}{c}  
   \includegraphics[height=8cm]{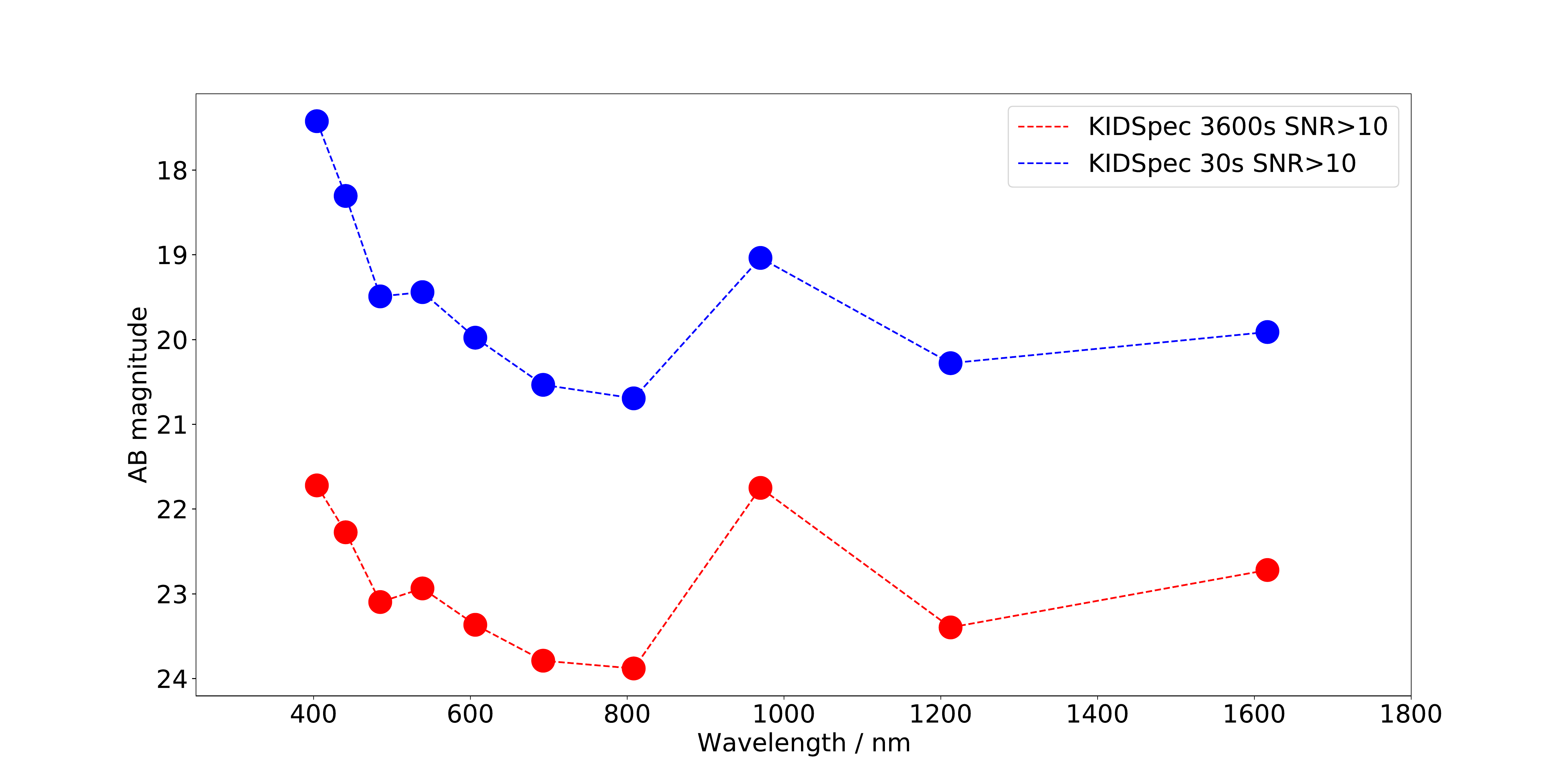}
   \end{tabular}
   \end{center}
   \caption[example] 
   { \label{fig:lim_mags} 
Limiting magnitudes for the 7500 spectral resolution design used in this work. Each grating order's blaze wavelength SNR was taken with a threshold of SNR$>$10. Two exposure times were considered, 3600s and 30s. The large peak at $\approx{1000}$nm is the result of an unfortunate bright sky line at this wavelength which reduced the limiting magnitude at this order's blaze wavelength. The 30s exposure limiting magnitudes at $\approx{400-500}$nm again approach the 19 magnitude of ZTFJ1539 + 5027, which would allow KIDSpec to take more phase bins of the period. }
   \end{figure}

The limiting magnitudes of this design of KIDSpec across its bandpass on the ELT were also found for an hour and 30s exposure, with a threshold of an SNR of 10. Each grating order's blaze wavelength SNR was measure here. Fig. \ref{fig:lim_mags} contains these results. The large peak at $\approx{1000}$nm is the result of an unfortunate bright sky line at this wavelength which reduced the limiting magnitude at this order's blaze wavelength. The 30s exposure limiting magnitudes at $\approx{400-500}$nm again approach the 19 magnitude of ZTFJ1539 + 5027, which would allow KIDSpec to take more phase bins of the period. For 3600s KIDSpec could reach a magnitude of $\approx{23.8}$ and $\approx{20.5}$ for 30s.

However the state of MKID research when the ELT construction is complete can also be considered, as this is still quite some time away, at least the second half of the 2020s\cite{Tamai2020TheProgress}. In 2013 ARCONS\cite{Mazin2013ARCONS:Spectrophotometer} had 2024 MKIDs at an average $R_{E}$ of 8. MEC\cite{Walter2020TheSCExAO} in 2020 had an average $R_{E}$ of 10 with 20,440 MKIDs. One of the focuses for MKID research in the OPT/NIR in the 2020s will be to improve the energy resolution of the detectors\cite{Mazin2019Optical2020s}. 



\section{Conclusions}

KIDSpec, an upcoming OPT/NIR MKID spectrograph, presents exciting opportunities for many science cases, discussed in this paper was the LISA verification sources which are binary systems with periods less than 30 minutes. One such system is ZTFJ1539 + 5027, a magnitude 19 system with a a period of 7 minutes. Typical CCD detector instruments struggle here with their read noise due to the short exposures required to constrain parameters of the system. A ZTFJ1539 + 5027 like system was simulated using KSIM and FORS' ETC. KIDSpec, using its read noise free MKIDs and microsecond time resolution will improve over these detectors, almost doubling the SNR of FORS for the same 10s observation simulation to 3.5 from 1.8.

LSST will search in photometry for these LISA verification sources and will able to detect systems with a magnitude brighter than $\approx{24}$. When extending this to simulate KIDSpec on an ELT class telescope for a night of observations, KIDSpec could observe systems with up to $m_{V}\approx{24.3}$ with an SNR>5. Here KIDSpec could build SNR in a particular orbital phase in the period of ZTFJ1539 + 5027 with a long continuous observation to reach these magnitudes. This work demonstrates KIDSpec's potential and flexibility for faint and short spectroscopy. KSIM can be obtained on request from the author.

\acknowledgments 
 
The authors would like to thank Dr. Tim Morris for helpful guidance and discussion on simulating the use of an ELT-like telescope. VBH and KOB are supported by the Science and Technology Facilities Council (STFC) under grant numbers ST/T506047/1 and ST/T002433/1 respectively. This project has received funding from the European Union's Horizon 2020 research and innovation programme under grant agreement No 730890.

\bibliography{main} 
\bibliographystyle{spiebib} 

\end{document}